\documentclass{PoS}
\usepackage{lineno}

\title{Dipion photoproduction and the $Q^2$ evolution of the shape of the gold nucleus}

\ShortTitle{Dipions and nuclear shape}

\author{\speaker{Spencer R. Klein for the STAR Collaboration}\\
       Lawrence Berkeley National Laboratory\\
       Berkeley CA 94720 USA \\
        E-mail: \email{srklein@lbl.gov}}

\abstract{Coherent photoproduction of vector mesons is sensitive to the shape of the target nucleus, as probed at four-momentum scale $Q^2 \propto (M_V/2)^2$.  Previously STAR presented a high-statistics measurement of $d\sigma_c/dt$ for coherent $\pi^+\pi^-$ photoproduction in ultra-peripheral gold-gold collisions, and made a two-dimensional Fourier-Bessel (Hanckel) transformation to give the transverse distribution of interactions in the nucleus.  Here, we study how $d\sigma_c/dt$ evolves with $Q^2$.  We divide the $\pi^+\pi^-$ signal into three different mass ($Q^2$) bins to measure how $d\sigma_c/dt$ evolves with dipion mass.  Furthermore, we find that the depth of the first diffractive minimum varies with pair mass.  We perform a two-dimensional Fourier-Bessel transform to see how the effective transverse distribution of the interactions changes with decreasing pair mass.  In the lowest mass bin, the nuclear profile is broader, which is consistent with expectations from saturation models.
}

\FullConference{XXVI International Workshop on Deep-Inelastic Scattering and Related Subjects (DIS2018)\\
		16-20 April 2018\\
		Kobe, Japan}

\begin{document}

\section{Introduction}

Nuclear shadowing reduces the cross-section for photon-nucleus interactions compared with a collection of the same number of
independent nucleons.   The reduction is often quantified as the change in cross-section for a quark-antiquark dipole (the quantum fluctuation of a photon).     In addition, shadowing affects other observables in the interaction.  This can be studied in the dipole picture, where an incident photon fluctuates to a quark-antiquark dipole, which then scatters elastically from the nucleus, emerging as a real vector meson. For small dipoles, with large $Q^2$ (squared 4-momentum transfer), the interactions can be described in terms of quarks and gluons.  At lower $Q^2$, as will be discussed here, individual quarks and gluons are not visible, but nuclear shadowing still affects the cross-section \cite{Bezrukov:1981ci,Armesto:2006ph}.  

Large dipoles tend to interact with the first nucleon that they encounter, on the front face of the nuclear target.  For sufficiently large dipoles, the nucleus appears as a black disk, with an equal interaction probability ($\approx 1$) at all impact parameters.    Small dipoles are less likely to interact, and thus can penetrate the nucleus.  They have an equal probability to interact with all of the nucleons, so the interaction sites follow the nuclear density profile.   Coherent interactions probe these two cases. For coherent elastic scattering (including vector meson photoproduction), one adds the amplitudes for scattering off of the individual nucleons $i$ at positions $\vec{x}_i$:
 \begin{equation}
 \sigma_c = \int  d^3\vec{k} |\Sigma_i A_i exp(i\vec{k}\cdot \vec{x}_i)|^2
 \end{equation}
where (neglecting the small virtuality) $\vec{k}$ is the momentum transfer from the nucleus to the nascent vector meson and $t=|\vec{k}|^2$.  We take the interaction amplitudes $A_i$ to be identical.   In high-energy photoproduction, the longitudinal component $k_z$ is small, and can be neglected.  This also avoids the  two-fold ambiguity in $k_z$ due to uncertainty in the photon direction.   By measuring $d\sigma_c/dt$  and performing a two-dimensional Fourier transform, we can learn about the positions where the dipoles interacted within the nucleus \cite{Diehl,Toll:2012mb,Adamczyk:2017vfu}.  

Here, we present a study of dipion  photoproduction in ultra-peripheral collisions (UPCs)~\cite{Bertulani:2005ru}.  In these UPCs, a photon emitted by one nucleus fluctuates to a quark-antiquark dipole which scatters in the nucleus and emerges as a dipion pair.  The final state may be produced via $\rho$ or $\omega$ intermediates, or by direct $\pi\pi$ photoproduction.  These three possibilities interfere, and the combination has been shown to provide a good description of the $\pi\pi$ mass spectrum in the range up to 1 GeV \cite{Adamczyk:2017vfu}.  The  final dipion $p_T$ comes mostly from the scattering ($k_T$), but also includes a component from the photon transverse momentum.

\section{STAR and data selection}

This analysis uses data taken with the Solenoidal Tracker at RHIC (STAR) detector \cite{Ackermann:2002ad} in 2010 and 2011, on gold-gold collisions at a center of mass energy of 200 GeV/nucleon pair.   For this analysis, the most important components of
STAR are the time projection chamber (TPC),  time-of-flight (TOF) system, zero-degree calorimeters (ZDCs) and forward beam-beam counters (BBCs).  The TPC  tracked charged particles with transverse momentum $p_T>200$ MeV/c and
pseudorapidity $|\eta|<1$.   The TOF system was used for triggering.  The trigger required 2-6 hits in the TOF system, with $|\eta|<1$, and signals equivalent to 1-4 neutrons in each ZDC.   The BBC was used to reject events with charged particles with $2 < |\eta| < 5$.

 The analysis used tight cuts to select photoproduced dipion pairs.  It selected events with exactly two tracks coming from the
 primary vertex.  The tracks had to be well reconstructed, with at  least 25 space points  in the TPC.   Track pairs with
 rapidity $|y|<0.03$ were rejected because they could be from cosmic-ray muons, which are reconstructed as a pair with $p_T=0$
 and $y=0$ .

This left 437K pairs, with the dipion invariant mass between 620 and 920 MeV.   The lower edge of the low-mass bin was chosen to largely remove background from two other UPC processes: $\gamma\gamma\rightarrow e^+e^-$ and $\gamma A \rightarrow \omega A$, followed by $\omega\rightarrow\pi^+\pi^-\pi^0$, where the $\pi^0$ is not seen by STAR.   This mass spectrum was divided into three
ranges with a similar number of events: 149K events (low), 148K events (medium) and 140K events (high mass).  In all three bins, the signal to noise ratio is more than $10:1$. 
 
\begin{figure}[t]
\includegraphics[width=.46\textwidth]{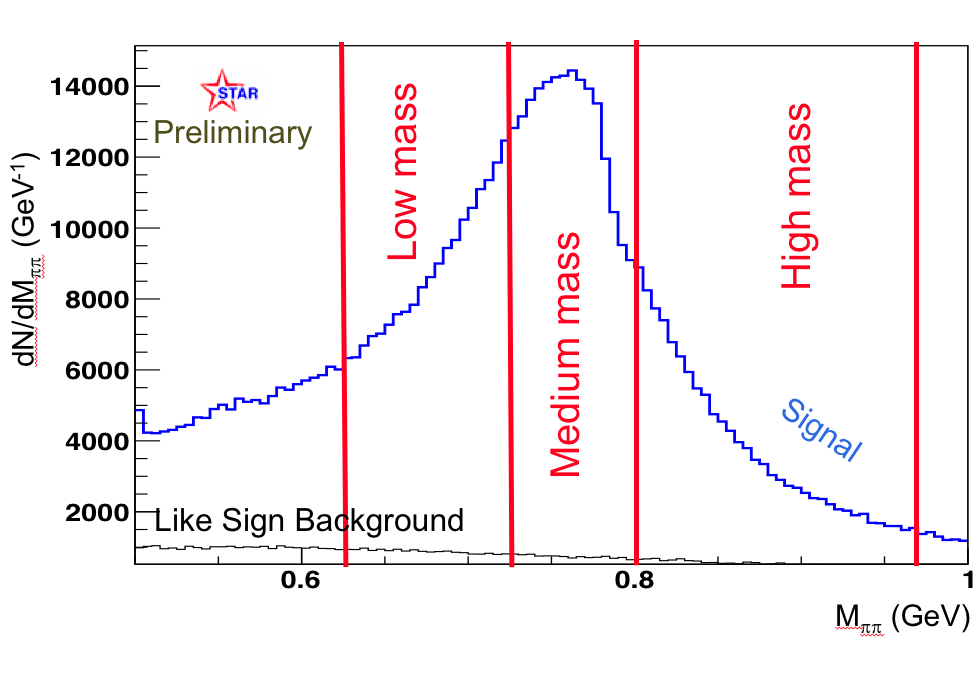}
\includegraphics[width=.46\textwidth]{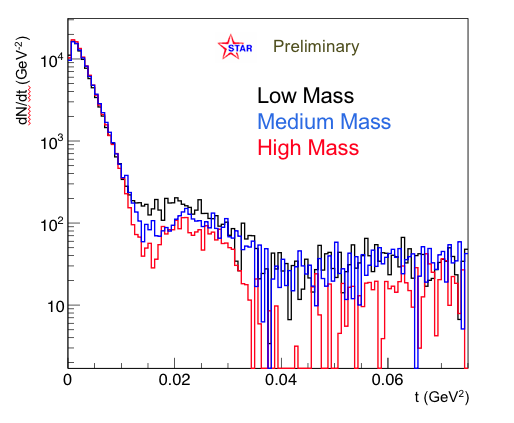}
 \caption{(left) The dipion mass spectrum, showing the three mass bins used in this analysis.  The blue curve shows the charge zero pairs, while the black shows the like-sign background.  (right)  $dN/dt$ for the three different mass ranges, after subtraction of the incoherent components.  The depths of the diffractive minima are  different.
 }
 \label{fig:mass}
 \end{figure}

\begin{figure}[t]
\includegraphics[width=.467\textwidth]{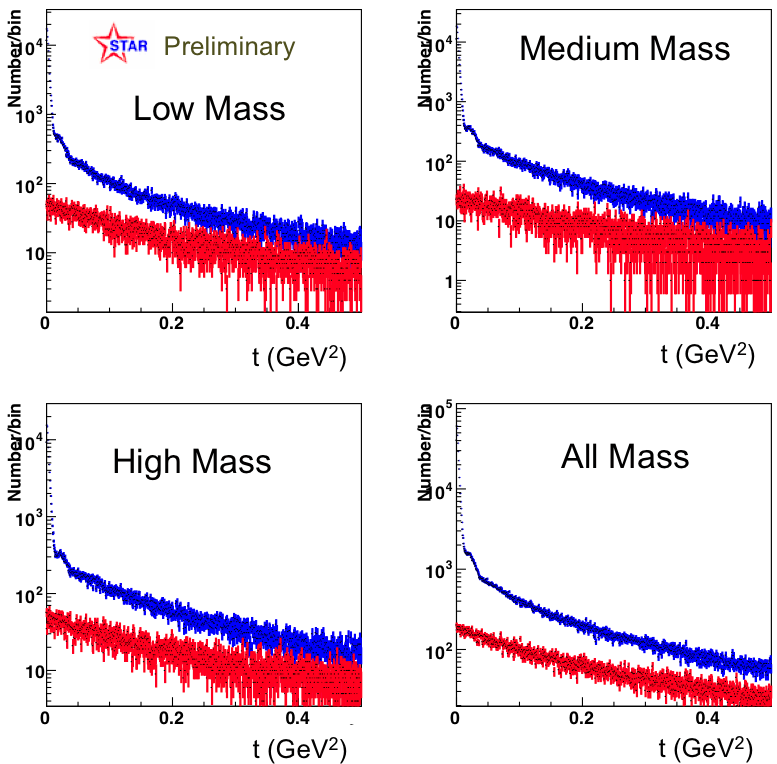}
\includegraphics[width=.48\textwidth]{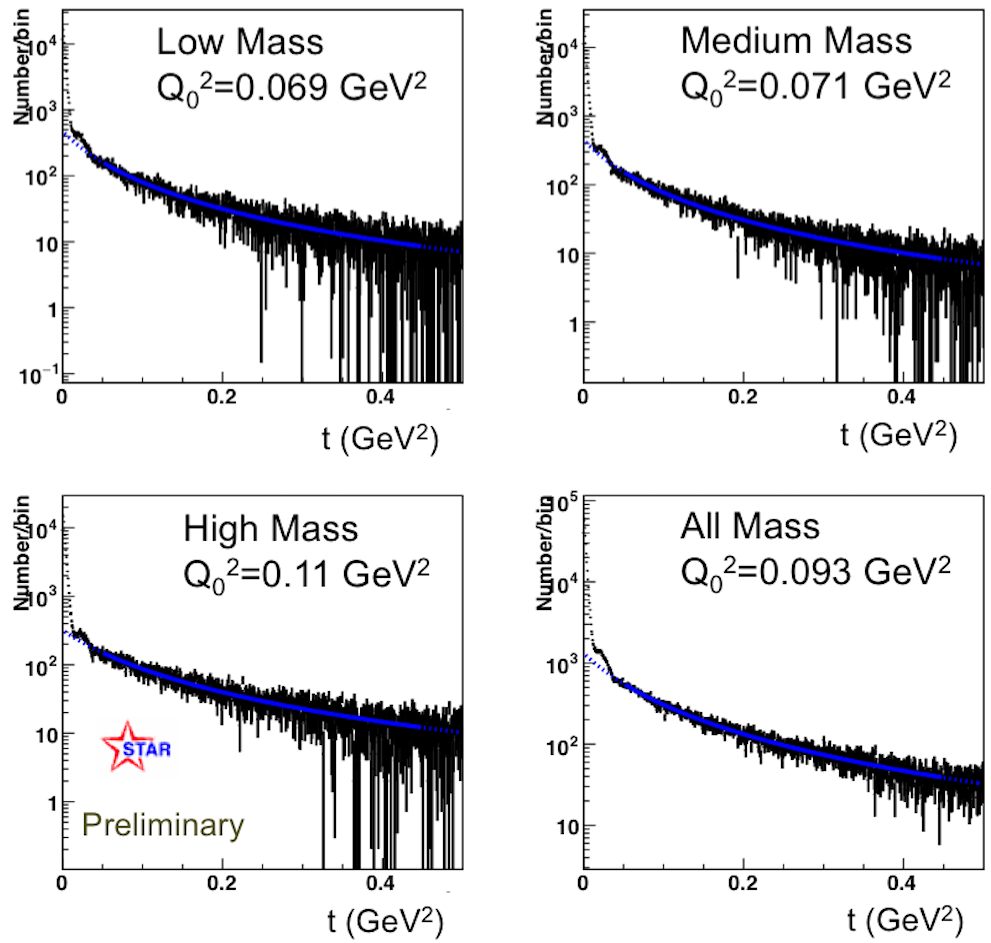}
 \caption{(left) The $dN/dt$ histograms for the three mass bins plus the summed bin.  The blue histograms are the net-charge-zero pairs, while the red shows the like-sign background.  The coherent production peak is visible for $t < 0.01$ GeV$^2$. (right) Like-sign subtracted total $d\sigma/dt$.  The blue line shows the dipole fit to the data, and the listed $Q_0$ are from that fit.}
 \label{fig:ptinit}
 \end{figure}

Figure \ref{fig:ptinit} (left) shows the $t$ spectra of the three mass ranges, along with the like-sign
background which is an estimate of the background from peripheral hadronic collisions.  Figure \ref{fig:ptinit} (right)  shows the $t$ distribution of the like-sign subtracted pairs.  To remove the incoherent background, we fit the histogram in the range $0.05\ {\rm GeV}^2 < t < 0.45$  GeV$^2$ to a dipole function:
\begin{equation}
\frac{dN} {dt} =\frac{A/Q_0^2}{(1+t/Q_0^2)^2}
\end{equation}
This is the same function used in Ref. \cite{Adamczyk:2017vfu}, but with a wider fit range.  All of the fits have $\chi^2/DOF\approx 1$.  
Ref \cite{Adamczyk:2017vfu} fixed $Q_0^2$ to be 0.099 GeV$^2$, while we let it float.  $Q_0^2$ shows some variation with mass bin, but the low-mass and medium-mass bins agree within their uncertainties.  Some variation of $Q_0^2$ with dipole size is not unexpected.

As Fig. \ref{fig:ptinit} (right)  shows, an exponential does not match the incoherent photoproduction data; it would be a straight line on the semi-log plot.  For the all-mass bin in the same $t$ range,  an exponential fit gives a $\chi^2/DOF =1345/639$, in contrast to the $\chi^2/DOF =
659/639$ for the dipole fit.  Although STAR previously used an exponential to model incoherent photoproduction \cite{Adler:2002sc,Abelev:2007nb}, with the greatly increased statistics, it is no longer a good fit to the data.

The incoherent contribution is then subtracted; the resulting $d\sigma_c/dt$ is shown in Fig. \ref{fig:mass} (right).   The depths of the first diffraction minima are different for the three curves. 

\section{The nuclear shape}

The two-dimensional nuclear shape profile, $F(b)$ can be given by
observing the two-dimensional Fourier-Bessel (Hanckel)  transform of $d\sigma_c/dt$
\cite{Diehl,Toll:2012mb}:
\begin{equation}
F(b) \propto \frac{1}{2\pi} \int_0^{\sqrt{t_{\rm max}}} dp_T p_T J_0(bp_T) \sqrt{\frac{d\sigma_c}{dt}}
\end{equation}
where $\sqrt{t_{ \rm max}}$ is the maximum $p_T$ and $J_0$ is a modified Bessel function.  There is one important caveat, due to the square root, which converts from cross-section to amplitude. 
Because the square root has two roots, one positive and one negative, one must flip the sign of $\sqrt{d\sigma_c/dt}$ at each diffractive minimum in $d\sigma_c/dt$.   

\begin{figure}[t]
\centerline{\includegraphics[width=.405\textwidth]{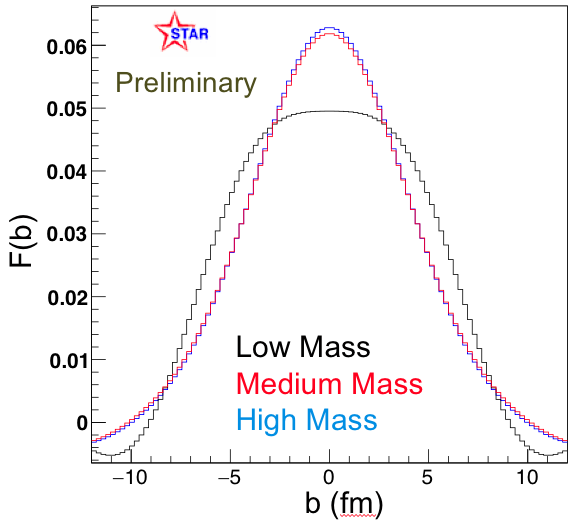}
\includegraphics[width=.4\textwidth]{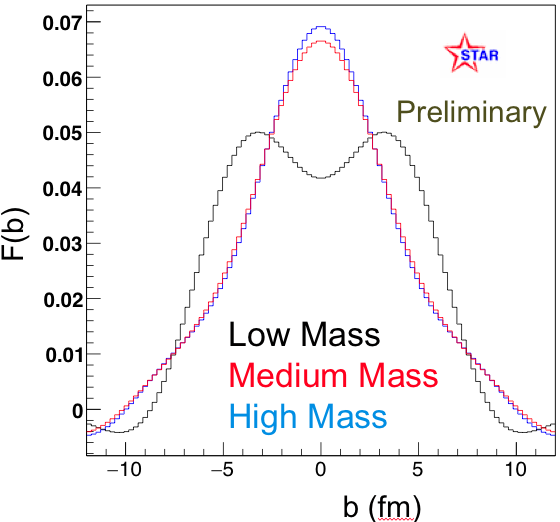}}
 \caption{Fourier transform $F(b)$ for the three different mass bins, with (left) $t_{\rm max} = 0.006$ GeV$^2$ and
 (right) $t_{\rm max} = 0.009$ GeV$^2$}
 \label{fig:fofb}
 \end{figure}

This relationship is exact for $t_{ \rm max}= \infty$.  Unfortunately, this is not experimentally accessible.  The imposition of a finite $t_{\rm max}$ can introduce artifacts into $F(b)$.  In signal-processing language, this is equivalent to the application of a square-pulse windowing function \cite{window}.   The output $F(b)$ is the product of the Fourier transform of the true nuclear shape with the Fourier transform of the windowing function.   There are techniques to moderate these artifacts by using a different windowing function, but they are problematic here because the window can only include 2-3 cycles (one cycle between each diffractive minimum).  So, we will not do this.  Instead, we will make transforms with different $t_{\rm max}$ and compare the results from the three  mass bins, for the same $t_{\rm max}$.  We use as a baseline $t_{\rm max} = 0.006$ GeV$^2$, which gives results compatible with the previous STAR  photoproduction paper.  It also avoids any of the sign flips due to the diffractive minima.  Figure~\ref{fig:fofb}~(left) shows $F(b)$ for the three mass ranges with this baseline.  The medium and high mass curves are similar, showing a peak and a smooth fall-off, while the low-mass curve is noticeably broader, with a flatter top.  This is expected if saturation is present; the profile will broaden, since absorption in the center cannot increase above one; in the black disk limit, $F(b)$ would be constant, with a sharp fall-off near the nuclear edges.  $F(b)$ drops below zero at large $|b|$; this is likely due to photoproduction in the opposing nucleus. Because of the parity inversion going from one nuclear target to the other, this inversion introduces a negative sign \cite{Abelev:2008ew}.

Figure \ref{fig:fofb} (right)  shows the effects of changing $t_{\rm max}$ from 0.006 to 0.009 GeV$^2$ for the three mass ranges.  The medium and high mass ranges do not change much, but the low-mass $F(b)$ grows considerably broader.  This 'double-hump' behavior is also seen in the original STAR data when $t_{\rm max}$ is increased \cite{Adamczyk:2017vfu}. We also studied a smaller $t_{\rm max}$, 0.005 GeV$^2$; it had a similar $F(b)$ to the baseline. 

\subsection{STARlight as a null experiment}

To understand the changes in $F(b)$ with $t_{\rm max}$, we apply a similar procedure to simulated dipion data from STARlight \cite{Klein:2016yzr}.    STARlight uses a Glauber calculation to determine the cross-section, but  it generates the $p_T$ from the photon-nucleus scattering using a simple model for gold nuclei, with the density following a Woods-Saxon distribution  \cite{Klein:1999qj}.  The photon $p_T$ are generated following the equivalent photon approximation \cite{Vidovic:1992ik}, including the interference between dipion production on the two nuclei \cite{Klein:1999gv}.   So, STARlight simulates the non-scattering aspects of the reaction that can affect $p_T$.   Five million STARlight $\rho^0 +$ direct $\pi\pi$ were generated within the kinematic range $|y|<1$, $|\eta_\pi| <1$ and $p_{T,\pi} > 100$ MeV/c, and divided up by mass range.    The $p_T$ spectra are indistinguishable, and the resulting $T(b)$ for all three mass bins  look similar to the high-mass data.   When $t_{\rm max}$ is varied, it also tends to follow the high-mass data.

\section{Conclusions and future work}

The diffractive minima in $d\sigma_c/dt$ show variation in depth with dipion mass.  The Fourier transform of $d\sigma_c/dt$, $F(b)$ also show apparent changes with pair mass.  In the lowest mass bin, $F(b)$ is flattened, consistent with the expectations from shadowing.  The transform is closer to a black disk than the other profiles, which are closer to the Woods-Saxon distribution that is implemented in STARlight.  

It is a pleasure to acknowledge useful conversations with Markus Diehl and Thomas Ullrich.  This work was funded by the U.S. DOE under contract number DE-AC-76SF0009.

\end{document}